\documentclass[aps,prd,twocolumn,showpacs,groupedaddress]{revtex4}  
\usepackage{graphicx}  
\usepackage{dcolumn}   
\usepackage{bm}        
\usepackage{amssymb}   
\begin{document}

\hspace{5.2in} \mbox{FERMILAB-PUB-06-426-E}

\title{Measurement of the $p\bar{p} \to t\bar{t}$ production cross section at $\sqrt{s}=1.96$~TeV in the fully hadronic decay channel }
%
\author{                                                                      
V.M.~Abazov,$^{35}$                                                           
B.~Abbott,$^{75}$                                                             
M.~Abolins,$^{65}$                                                            
B.S.~Acharya,$^{28}$                                                          
M.~Adams,$^{51}$                                                              
T.~Adams,$^{49}$                                                              
E.~Aguilo,$^{5}$                                                              
S.H.~Ahn,$^{30}$                                                              
M.~Ahsan,$^{59}$                                                              
G.D.~Alexeev,$^{35}$
% ...
\\                                                                            
\vskip 0.30cm                                                                 
\centerline{(D\O\ Collaboration)}                                             
\vskip 0.30cm                                                                 
}                                                                             
\affiliation{                                                                 
\centerline{$^{1}$Universidad de Buenos Aires, Buenos Aires, Argentina}       
\centerline{$^{2}$LAFEX, Centro Brasileiro de Pesquisas F{\'\i}sicas,         
                  Rio de Janeiro, Brazil}                                     
\centerline{$^{3}$Universidade do Estado do Rio de Janeiro,                   
                  Rio de Janeiro, Brazil}                                     
\centerline{$^{4}$Instituto de F\'{\i}sica Te\'orica, Universidade            
                  Estadual Paulista, S\~ao Paulo, Brazil}                     
\centerline{$^{5}$University of Alberta, Edmonton, Alberta, Canada,           
                  Simon Fraser University, Burnaby, British Columbia, Canada,}
\centerline{York University, Toronto, Ontario, Canada, and                    
                  McGill University, Montreal, Quebec, Canada}                
\centerline{$^{6}$University of Science and Technology of China, Hefei,       
                  People's Republic of China}                                 
\centerline{$^{7}$Universidad de los Andes, Bogot\'{a}, Colombia}             
\centerline{$^{8}$Center for Particle Physics, Charles University,            
                  Prague, Czech Republic}                                     
\centerline{$^{9}$Czech Technical University, Prague, Czech Republic}         
\centerline{$^{10}$Center for Particle Physics, Institute of Physics,         
                   Academy of Sciences of the Czech Republic,                 
                   Prague, Czech Republic}                                    
\centerline{$^{11}$Universidad San Francisco de Quito, Quito, Ecuador}        
\centerline{$^{12}$Laboratoire de Physique Corpusculaire, IN2P3-CNRS,         
                   Universit\'e Blaise Pascal, Clermont-Ferrand, France}      
\centerline{$^{13}$Laboratoire de Physique Subatomique et de Cosmologie,      
                   IN2P3-CNRS, Universite de Grenoble 1, Grenoble, France}    
\centerline{$^{14}$CPPM, IN2P3-CNRS, Universit\'e de la M\'editerran\'ee,     
                   Marseille, France}                                         
\centerline{$^{15}$Laboratoire de l'Acc\'el\'erateur Lin\'eaire,              
                   IN2P3-CNRS et Universit\'e Paris-Sud, Orsay, France}       
\centerline{$^{16}$LPNHE, IN2P3-CNRS, Universit\'es Paris VI and VII,         
                   Paris, France}                                             
\centerline{$^{17}$DAPNIA/Service de Physique des Particules, CEA, Saclay,    
                   France}                                                    
\centerline{$^{18}$IPHC, IN2P3-CNRS, Universit\'e Louis Pasteur, Strasbourg,  
                   France, and Universit\'e de Haute Alsace,                  
                   Mulhouse, France}                                          
\centerline{$^{19}$Institut de Physique Nucl\'eaire de Lyon, IN2P3-CNRS,      
                   Universit\'e Claude Bernard, Villeurbanne, France}         
\centerline{$^{20}$III. Physikalisches Institut A, RWTH Aachen,               
                   Aachen, Germany}                                           
\centerline{$^{21}$Physikalisches Institut, Universit{\"a}t Bonn,             
                   Bonn, Germany}                                             
\centerline{$^{22}$Physikalisches Institut, Universit{\"a}t Freiburg,         
                   Freiburg, Germany}                                         
\centerline{$^{23}$Institut f{\"u}r Physik, Universit{\"a}t Mainz,            
                   Mainz, Germany}                                            
\centerline{$^{24}$Ludwig-Maximilians-Universit{\"a}t M{\"u}nchen,            
                   M{\"u}nchen, Germany}                                      
\centerline{$^{25}$Fachbereich Physik, University of Wuppertal,               
                   Wuppertal, Germany}                                        
\centerline{$^{26}$Panjab University, Chandigarh, India}                      
\centerline{$^{27}$Delhi University, Delhi, India}                            
\centerline{$^{28}$Tata Institute of Fundamental Research, Mumbai, India}     
\centerline{$^{29}$University College Dublin, Dublin, Ireland}                
\centerline{$^{30}$Korea Detector Laboratory, Korea University,               
                   Seoul, Korea}                                              
\centerline{$^{31}$SungKyunKwan University, Suwon, Korea}                     
\centerline{$^{32}$CINVESTAV, Mexico City, Mexico}                            
\centerline{$^{33}$FOM-Institute NIKHEF and University of                     
                   Amsterdam/NIKHEF, Amsterdam, The Netherlands}              
\centerline{$^{34}$Radboud University Nijmegen/NIKHEF, Nijmegen, The          
                  Netherlands}                                                
\centerline{$^{35}$Joint Institute for Nuclear Research, Dubna, Russia}       
\centerline{$^{36}$Institute for Theoretical and Experimental Physics,        
                   Moscow, Russia}                                            
\centerline{$^{37}$Moscow State University, Moscow, Russia}                   
\centerline{$^{38}$Institute for High Energy Physics, Protvino, Russia}       
\centerline{$^{39}$Petersburg Nuclear Physics Institute,                      
                   St. Petersburg, Russia}                                    
\centerline{$^{40}$Lund University, Lund, Sweden, Royal Institute of          
                   Technology and Stockholm University, Stockholm,            
                   Sweden, and}                                               
\centerline{Uppsala University, Uppsala, Sweden}                              
\centerline{$^{41}$Physik Institut der Universit{\"a}t Z{\"u}rich,            
                   Z{\"u}rich, Switzerland}                                   
\centerline{$^{42}$Lancaster University, Lancaster, United Kingdom}           
\centerline{$^{43}$Imperial College, London, United Kingdom}                  
\centerline{$^{44}$University of Manchester, Manchester, United Kingdom}      
\centerline{$^{45}$University of Arizona, Tucson, Arizona 85721, USA}         
\centerline{$^{46}$Lawrence Berkeley National Laboratory and University of    
                   California, Berkeley, California 94720, USA}               
\centerline{$^{47}$California State University, Fresno, California 93740, USA}
\centerline{$^{48}$University of California, Riverside, California 92521, USA}
\centerline{$^{49}$Florida State University, Tallahassee, Florida 32306, USA} 
\centerline{$^{50}$Fermi National Accelerator Laboratory,                     
            Batavia, Illinois 60510, USA}                                     
\centerline{$^{51}$University of Illinois at Chicago,                         
            Chicago, Illinois 60607, USA}                                     
\centerline{$^{52}$Northern Illinois University, DeKalb, Illinois 60115, USA} 
\centerline{$^{53}$Northwestern University, Evanston, Illinois 60208, USA}    
\centerline{$^{54}$Indiana University, Bloomington, Indiana 47405, USA}       
\centerline{$^{55}$University of Notre Dame, Notre Dame, Indiana 46556, USA}  
\centerline{$^{56}$Purdue University Calumet, Hammond, Indiana 46323, USA}    
\centerline{$^{57}$Iowa State University, Ames, Iowa 50011, USA}              
\centerline{$^{58}$University of Kansas, Lawrence, Kansas 66045, USA}         
\centerline{$^{59}$Kansas State University, Manhattan, Kansas 66506, USA}     
\centerline{$^{60}$Louisiana Tech University, Ruston, Louisiana 71272, USA}   
\centerline{$^{61}$University of Maryland, College Park, Maryland 20742, USA} 
\centerline{$^{62}$Boston University, Boston, Massachusetts 02215, USA}       
\centerline{$^{63}$Northeastern University, Boston, Massachusetts 02115, USA} 
\centerline{$^{64}$University of Michigan, Ann Arbor, Michigan 48109, USA}    
\centerline{$^{65}$Michigan State University,                                 
            East Lansing, Michigan 48824, USA}                                
\centerline{$^{66}$University of Mississippi,                                 
            University, Mississippi 38677, USA}                               
\centerline{$^{67}$University of Nebraska, Lincoln, Nebraska 68588, USA}      
\centerline{$^{68}$Princeton University, Princeton, New Jersey 08544, USA}    
\centerline{$^{69}$State University of New York, Buffalo, New York 14260, USA}
\centerline{$^{70}$Columbia University, New York, New York 10027, USA}        
\centerline{$^{71}$University of Rochester, Rochester, New York 14627, USA}   
\centerline{$^{72}$State University of New York,                              
            Stony Brook, New York 11794, USA}                                 
\centerline{$^{73}$Brookhaven National Laboratory, Upton, New York 11973, USA}
\centerline{$^{74}$Langston University, Langston, Oklahoma 73050, USA}        
\centerline{$^{75}$University of Oklahoma, Norman, Oklahoma 73019, USA}       
\centerline{$^{76}$Oklahoma State University, Stillwater, Oklahoma 74078, USA}
\centerline{$^{77}$Brown University, Providence, Rhode Island 02912, USA}     
\centerline{$^{78}$University of Texas, Arlington, Texas 76019, USA}          
\centerline{$^{79}$Southern Methodist University, Dallas, Texas 75275, USA}   
\centerline{$^{80}$Rice University, Houston, Texas 77005, USA}                
\centerline{$^{81}$University of Virginia, Charlottesville,                   
            Virginia 22901, USA}                                              
\centerline{$^{82}$University of Washington, Seattle, Washington 98195, USA}  
}                                                                             
\date{December 13, 2006}

\begin{abstract}
A measurement of the top quark pair production cross section in proton anti-proton collisions at an interaction energy of $\sqrt{s}=1.96~{\rm TeV}$ is presented. This analysis uses 405~pb$^{-1}$ of data collected with the D\O\ detector at the Fermilab Tevatron Collider. 
Fully hadronic $t\bar{t}$ decays with final states of six or more jets are separated from the multijet background using secondary vertex tagging and a neural network.
The $t\bar{t}$ cross section is measured as $\sigma_{t\bar{t}}=4.5_{-1.9}^{+2.0}({\rm stat}) _{-1.1}^{+1.4}({\rm syst}) \pm 0.3 ({\rm lumi})~{\rm pb }$ for a top quark mass of $m_{t} = 175~{\rm GeV/c^2}$.
\end{abstract}

\maketitle 

The standard model (SM) predicts that the top quark decays primarily into a $W$ boson and a $b$ quark. 
The measurement presented here tests the prediction of the SM in the dominant decay mode of the $t\bar{t}$ system: when both $W$ bosons decay to quarks, the so-called fully hadronic decay channel. 
This topology occurs in 46\% of $t\bar{t}$ events.  
The theoretical signature for fully hadronic $t\bar{t}$ events is six or more jets originating from the hadronization of the six quarks. 
Of the six jets, two originate from $b$ quark decays.  
Fully hadronic $t\bar{t}$ events are difficult to identify at hadron colliders because the background rate is many orders of magnitude larger than that of the $t\bar{t}$ signal.

We report a measurement of the production cross-section of top quark pairs, $\sigma_{t\bar{t}}$, using data collected with D\O\ in the fully hadronic channel,
that exploits the long lifetime of the $b$ hadrons in identifying $b$ jets. 
To increase the sensitivity for $t\bar{t}$ events, we used a neural network to distinguish signal from the overwhelming background of multijet production through Quantum Chromodynamic processes (QCD).

The D\O\ detector\ \cite{d0det} has a central tracking system consisting of a silicon micro strip tracker (SMT) and a central fiber tracker (CFT), both located within a 2~T superconducting solenoidal magnet, with designs optimized for tracking and vertexing at pseudorapidities $|\eta|<3$ and $|\eta|<2.5$, respectively. 
Rapidity $y$ and pseudorapidity $\eta$ are defined as functions of the polar angle $\theta$ and parameter $\beta$ as $y(\theta,\beta)= \frac{1}{2} \ln [ (1+\beta \cos \theta)/(1-\beta \cos \theta )]$ and $\eta(\theta)=y(\theta,1)$, where $\beta$ is the ratio of the particle's momentum to its energy. 
The liquid-argon and uranium calorimeter has a central section (CC) covering pseudorapidities $|\eta|$ up to $\approx 1.1$ and two end calorimeters (EC) that extend coverage to $|\eta| \approx 4.2$, with all three housed in separate cryostats. 
Each calorimeter cryostat contains a multilayer electromagnetic calorimeter, a finely segmented hadronic calorimeter and a third hadronic calorimeter that is more coarsely segmented, providing both segmentation in depth and in projective towers of size $0.1 \times 0.1$ in $\eta$-$\phi$ space, where $\phi$ is the azimuthal angle in radians.
An outer muon system, covering $|\eta|<2$, consists of a layer of tracking detectors and scintillation trigger counters in front of 1.8~T iron toroids, followed by two similar layers after the toroids. 
The luminosity is measured using plastic scintillator arrays placed in front of the EC cryostats. 

The data set was collected between 2002 and 2004, and corresponds to an integrated luminosity $\mathcal{L}=405~\pm~25~{\rm pb}^{-1}$ \cite{newlumi}. 
To isolate events with six jets, we used a dedicated multijet trigger. 
The requirements on the trigger, particularly on jet and trigger tower energy thresholds, were tightened during the collection of the data set to manage the increasing instantaneous luminosities delivered by the Fermilab Tevatron Collider. 
The change in trigger requirements had little effect on the efficiency for signal, while removing an increasing number of background events\ \cite{footnote1}. 
The trigger was tuned for the fully hadronic $t\bar{t}$ channel and was optimized to remain as efficient possible while using limited bandwidth.
The collection rate after all trigger levels was fixed to a few Hz, which was completely dominated by QCD multijet events as the hadronic $t\bar{t}$ event production rate is expected to be a few events per day. 
We required three or four trigger towers above an energy threshold of 5 GeV at the first trigger level, three reconstructed jets with transverse energies ($E_T$) above 8 GeV at the second trigger level, combined with a requirement on the sum of the transverse momenta ($p_{T}$) of the jets, and four or five reconstructed jets at transverse energy thresholds between 10 and 30 GeV at the highest trigger level\ \cite{d0det}.

We simulated $t\bar{t}$ production using {\sc alpgen 1.3} to generate the parton-level processes, and {\sc pythia 6.2} to model hadronization\ \cite{alpgen,pythia}. 
We used a top quark invariant mass of $m_{t}=175~{\rm GeV/c^2}$.
The decay of hadrons carrying bottom quarks was modeled using {\sc evtgen}\ \cite{evtgen}. 
The simulated $t\bar{t}$ events were processed with the full {\sc geant}-based D\O\ detector simulation, after which the Monte Carlo (MC) events were passed through the same reconstruction program as was used for data. 
The small differences between the MC model and the data were corrected by matching the properties of the reconstructed objects. 
The residual differences were very small and were corrected using factors derived from detailed
comparisons between the MC model and the data for well understood SM processes 
such as the jets in $Z$ boson and QCD dijet production.
 
In the offline analysis, jets were defined with an iterative cone algorithm\ \cite{jetsdef}. 
Before the jet algorithm was applied, calorimeter noise was suppressed by 
removing isolated cells whose measured energy was lower than four standard deviations above cell pedestal.  
In the case that a cell above this threshold was found to be adjacent to one with an energy less than four standard deviations above pedestal, the latter was retained if its signal exceeded 2.5 standard deviations above pedestal.
Cells that were reconstructed with negative energies were always removed. 


The elements for cone jet reconstruction consisted of projective towers of calorimeter cells. 
First, seeds were defined using a preclustering algorithm, using calorimeter towers above an energy threshold of 0.5 GeV. 
The cone jet reconstruction, an iterative clustering process where the jet axis was required to match the axis of a projective cone, was then run using all preclusters above 1.0 GeV as seeds. 
As jets from $t\bar{t}$ production are relatively narrow due to relatively high jet $p_{T}$, the jets were defined using a cone with radius $R_{{\rm cone}}=0.5$, where $\Delta R = \sqrt{(\Delta y)^2+(\Delta \phi)^2}$ . 
The resulting jets (proto-jets) took into account all energy deposits contained in the jet cone. 
If two proto-jets were within $1<\Delta R / R_{{\rm cone}} <2$, an additional midpoint clustering was applied, where the combination of the two proto-jets was used as a seed for a possible additional proto-jet. 
At this stage, the proto-jets that share transverse momentum were examined with a splitting and merging algorithm, after which each calorimeter tower was assigned to one proto-jet at most. The proto-jets were merged if the shared $p_T$ exceeded 50\% of the $p_T$ of the proto-jet with the lowest transverse momentum and the towers were added to the most energetic proto-jet while the other candidate was rejected. 
If the proto-jets shared less than half of their $p_{T}$, the shared towers were assigned to the proto-jet which was closest in $\Delta R$ space.
The collection of stable proto-jets remaining was then referred to as the {\em reconstructed} jets in the event. 
The minimal $p_{T}$ of a reconstructed jet was required to be 8~GeV/$c$ before any energy corrections were applied.

We removed jets caused by electromagnetic particles and jets resulting from noise in hadronic sections of the calorimeter by requiring that the fraction of the jet energy deposited in the calorimeter ($EMF$) was $0.05 < EMF < 0.95$ and the fraction of energy in the coarse hadronic calorimeter was less than 0.4. 
Jets formed from clusters of calorimeter cells known to be affected by noise were also rejected. The remaining noise contribution was removed by requiring that the jet also fired the first level trigger. 

To correct the calorimeter jet energies back to the level of particle jets, a jet energy scale (JES) correction $C^{JES}$ was applied. 
The same procedure has to be applied to Monte Carlo jets to ensure an identical calorimeter response in data and simulation. 
The particle level or true jet energy $E^{true}$ was obtained from the measured jet energy $E^m$ and the detector pseudorapidity, measured from the center of the detector ($\eta_{det}$), using the relation
\begin{equation}
E^{true}= \frac{E^m - E_0 ( \eta_{det}, \mathcal{L})}{\mathcal{R}(\eta_{det}, E^m) S(\eta_{det},E^m)} = C^{JES} (E^m,\eta_{det},\mathcal{L}) \cdot E^m .
\end{equation}
In data and MC the total correction was applied to the measured energy $E^m$ as a multiplicative factor $C^{JES}$. 
$E_{0}(\eta_{det},\mathcal{L})$ was the offset energy created by electronics noise and noise signal caused by the uranium in the calorimeter, pile-up energy from previous collisions and the additional energy from the underlying physics event. 
The dependence on the luminosity $\mathcal{L}$ was caused by the fact that the number of additional interactions was dependent on the instantaneous luminosity, while the dependence on $y$ was caused by variations in the calorimeter occupancy as a function of the jet rapidity. $\mathcal{R}(\eta_{det},E^m)$ parameterized the energy response of the calorimeter, while $S(\eta_{det},E^m)$ represents the fraction of the true partonic jet energy that was deposited inside the jet cone. 
This out-of-cone showering correction depended on the energy of the jet and its location in the calorimeter.

The JES was measured directly using $p_T$ conservation in photon + jet events. The method was identical for data and simulation and used transverse momentum balancing between the jet and the photon. 
As the energy scale of the photon was directly and precisely measured (the electromagnetic calorimeter response was derived from measurements of resonances in the $e^+ e^-$ spectrum like the $Z$ boson), the true jet energy could be derived from the difference between the photon and jet energy. $E_0$, $\mathcal{R}$ and $S$ were fit as a function of jet rapidity and measured energy, which lead to uncertainties coming from the fit (statistical) and the method (systematic). The total correction $C^{JES}$ was approximately 1.4 for data jets in the energy range expected for jets associated with top quark events. 
The uncertainties on $C^{JES}$, which were dominated by the systematic uncertainty of the out-of-cone showering correction $S(\eta_{det},E^m)$, were a few percent and were dependent on the jet energy and rapidity.

The jet energy resolution was measured in photon + jet data for low jet energies and dijet data for higher jet energy values. 
Fits to the transverse energy asymmetry $[p_T(1) - p_T(2)]/[p_T(1)+p_T(2)]$ between the transverse momenta of the  back-to-back jets and/or photon ($p_T(1)$ and $p_T(2)$) were then used to obtain the jet energy resolution as a function of jet rapidity and transverse energy. 
The uncertainties on the jet energy resolution were dominated by limited statistics in the samples used. 

In this analysis, we considered a data set consisting of events with four or more reconstructed jets, in which the scalar
sum of the uncorrected transverse momenta $H_T^{uncorr}$ of all the jets in the event
was greater than 90 GeV/$c$. 
The final analysis sample was a subset of this sample, where at least six jets with corrected 
transverse momentum greater than 15 GeV/$c$ and $|y|<2.5$ were required. 
Events with isolated high transverse momentum electron or muon candidates were vetoed to 
ensure that the all-hadronic and leptonic $t\bar{t}$ samples were disjoint\ \cite{ttbarlepjetsrun2,ttbarlepjetsvtxtag}.
In addition, we rejected events where two distinct $p\bar{p}$ interactions with separate primary vertices were observed and the jets in the event were not assigned to only one of the two primary vertices. 
The primary vertex requirement did not affect minimum bias interactions or $t\bar{t}$ events. 
Table\ \ref{effmostcuts} lists the efficiencies after the first set of selection cuts, commonly referred to as preselection, which includes the requirements on the primary vertex, the number of reconstructed jets and the presence of isolated leptons, and the efficiency after preselection and after preselection and the trigger.
Besides selecting all hadronic $t\bar{t}$ events, the analysis was also expected to accept a small contribution from the semi-leptonic (lepton+jets) $t\bar{t}$ decay channel. 
The combined efficiency included the fully hadronic and semi-leptonic $W$-boson branching fractions of $0.4619\pm0.0048$ and $0.4349\pm0.0027$ respectively\ \cite{pdg}.

We used a secondary vertex tagging algorithm (SVT) to identify $b$-quark jets. 
The algorithm was the same as used in previously published D\O\ $t\bar{t}$ production cross section measurements\ \cite{ttbarlepjetsrun2,ttbarlepjetsvtxtag}.
Secondary vertex candidates were reconstructed from two or more tracks in the jet, removing vertices consistent with originating from long-lived light hadrons as for example $K_S^0$ and $\Lambda$.
 Two configurations of the secondary vertex algorithm were used; these were labeled ``loose'' and ``tight'' respectively. 
If a reconstructed secondary vertex in the jet had a transverse decay length $L_{xy}$ significance ($L_{xy}/\sigma_{L_{xy}}$) $>$ 5 (7), the jet was tagged as a loose (tight) $b$-quark jet. 
The loose SVT was chosen to efficiently identify $b$-quark jets, while the tight SVT was configured to accept only very few light quark jets while sacrificing a small reduction in the efficiency for $b$-quark jets. 
Events with two or more loosely tagged jets were called double-tag events. 
The sample that did not contain two loosely tagged jets was inspected for events with one tight tag.  
Events thus isolated were labeled single-tag events.
The fully exclusive samples of single-tag and double-tag events were treated separately because of their different signal-to-background ratios. 
The use of the tight SVT selection for single tagged events optimized the rejection of mistags, the main background in the single-tag analysis.
When two tags were required, the background sample started to be dominated by direct $b\bar{b}$ production. 
The choice to use the loose SVT optimized the double-tag analysis for signal efficiency instead of background rejection.

\begin{table}[bt]
\caption{\label{effmostcuts} Efficiency for selection criteria applied before $b$-jet identification. 
Efficiencies listed include the efficiency for all previous selection criteria. The trigger efficiency is quoted for events that have passed the preselection. 
The uncertainties are due to Monte Carlo statistics. Listed are the selection efficiencies as determined for $t\bar{t}$ in the hadronic decay channel, the lepton+jets decay channel and the efficiency for all different decay channels corrected for $W$ boson branching fractions.
}
\begin{tabular}{l c c c}
\hline \hline
cut & $t\bar{t} \to {\rm hadrons}$ & $t\bar{t} \to \ell + {\rm jets}$ & any $t\bar{t}$ \\ \hline
{\small preselection} & {\small$0.2706 \pm 0.0016$} & {\small$0.0311\pm0.0008$} & {\small$0.1385\pm 0.0011$} \\
{\small trigger} & {\small$0.2527\pm 0.0015$}& {\small$0.0268 \pm 0.0007$} & {\small$0.1284 \pm 0.0010$}\\
\hline \hline
\end{tabular}
\end{table}

Compared to light-quark QCD multijet events, $t\bar{t}$ events on average have more jets of higher energy and with less boost in the beam direction, resulting in events with many central jets that all have similar and relatively high energies. 
Moreover, the fully hadronic decay makes it possible to reconstruct the $W$ boson and $t$ quark four-momenta. 
To distinguish between signal and background, we used the following event characteristics\ \cite{run1alljetsxsec1}:

\begin{figure}
\includegraphics[width=\linewidth]{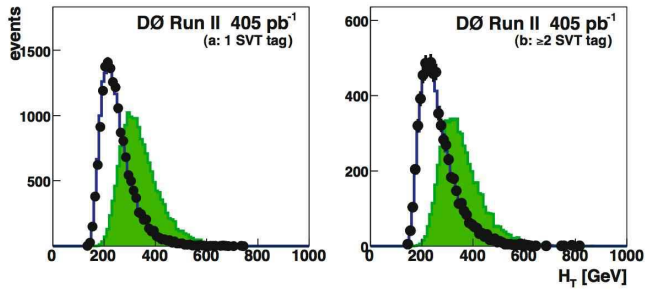}
\caption{\label{fig1} The $H_T$ distribution for single-tag events (a) and double-tag events (b). Shown are the data (points), the background (solid line) and the expected $t\bar{t}$ distribution (filled histogram) multiplied by 140 (60) for the single (double)-tag analysis.}
\end{figure}
(1) $H_T$: The scalar sum of the corrected transverse momenta of the jets (Fig.\ \ref{fig1}). 

\begin{figure}
\includegraphics[width=\linewidth]{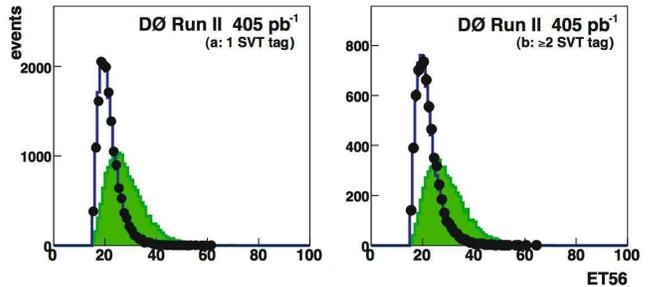}
\caption{\label{fig2} The $E_{T}^{56}$ distribution for single-tag events (a) and double-tag events (b). Shown are the data (points), the background (solid line) and the expected $t\bar{t}$ distribution (filled histogram) multiplied by 140 (60) for the single (double)-tag analysis.}
\end{figure}

(2) $E_{T}^{56}$: The square root of the product of the transverse momenta of the fifth and sixth leading jet (Fig.\ \ref{fig2}).

\begin{figure}
\includegraphics[width=\linewidth]{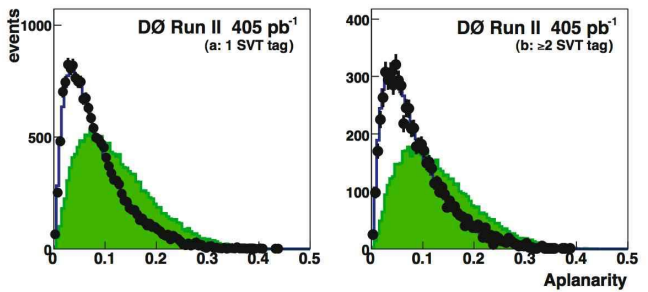}
\caption{\label{fig3} The $\mathcal{A}$ distribution for single-tag events (a) and double-tag events (b). Shown are the data (points), the background (solid line) and the expected $t\bar{t}$ distribution (filled histogram) multiplied by 140 (60) for the single (double)-tag analysis.}
\end{figure}

(3) $\mathcal{A}$: The aplanarity as calculated from the normalized momentum tensor (Fig.\ \ref{fig3})\ \cite{ttbarlepjetsrun2,ttbarlepjetsvtxtag,run1alljetsxsec1}. 

\begin{figure}
\includegraphics[width=\linewidth]{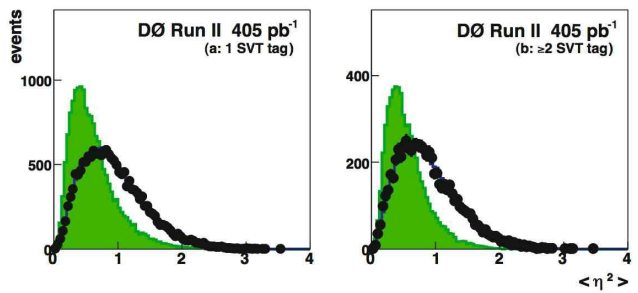}
\caption{\label{fig4} The $\langle \eta^2 \rangle$ distribution for single-tag events (a) and double-tag events (b). Shown are the data (points), the background (solid line) and the expected $t\bar{t}$ distribution (filled histogram) multiplied by 140 (60) for the single (double)-tag analysis.}
\end{figure}

(4) $\langle \eta^2 \rangle$: The $p_T$-weighted mean square of the $y$ of the jets in an event (Fig.\ \ref{fig4}), see also Ref.\  \cite{run1alljetsxsec1}. 

\begin{figure}
\includegraphics[width=\linewidth]{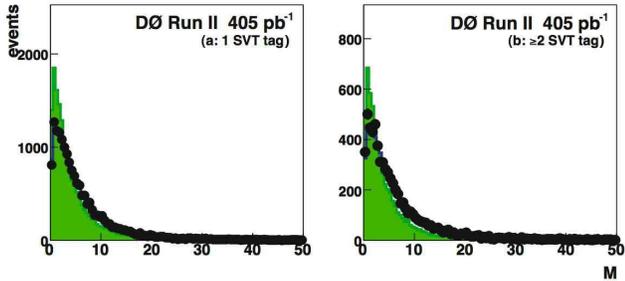}
\caption{\label{fig5} The $\mathcal{M}$ distribution for single-tag events (a) and double-tag events (b). Shown are the data (points), the background (solid line) and the expected $t\bar{t}$ distribution (filled histogram) multiplied by 140 (60) for the single (double)-tag analysis.}
\end{figure}

(5) $\mathcal{M}$: The mass-$\chi^2$ variable, which was defined as $\mathcal{M} = (M_{W_1}- M_W)^2/\sigma_{M_W}^2 +(M_{W_2}- M_W)^2/\sigma_{M_W}^2 +(m_{t_1}-m_{t_2})^2/\sigma_{m_t}^2$, where the parameters $M_W$, $\sigma_{M_W}$ and $\sigma_{m_t}$ were the invariant mass and mass resolution from the jet four-momenta calculated as observed in all-hadronic $t\bar{t}$ MC, respectively 79, 11 and 21 GeV/$c^2$ after all corrections and resolutions were included\ \cite{footnote2}.
$M_{W_i}$ and $m_{t_i}$ were calculated for every possible permutation of the jets in the event. We did not distinguish between tagged and untagged jets.  The combination of jets that yielded the lowest value of $\mathcal{M}$ is used (Fig.\ \ref{fig5}).

\begin{figure}
\includegraphics[width=\linewidth]{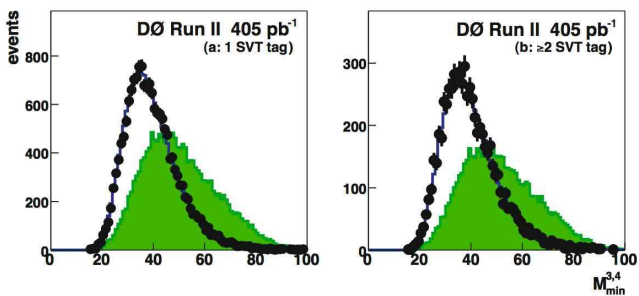}
\caption{\label{fig6} The $M_{min}^{34}$ distribution for single-tag events (a) and double-tag events (b). Shown are the data (points), the background (solid line) and the expected $t\bar{t}$ distribution (filled histogram) multiplied by 140 (60) for the single (double)-tag analysis.}
\end{figure}

(6) $M_{min}^{34}$: The second-smallest dijet mass in the event. First, all possible dijet masses were considered and the jets that yield the smallest mass were rejected. $M_{min}^{34}$ was the smallest dijet mass as found from the remaining jets (Fig.\ \ref{fig6}).

\begin{figure}
\includegraphics[width=\linewidth]{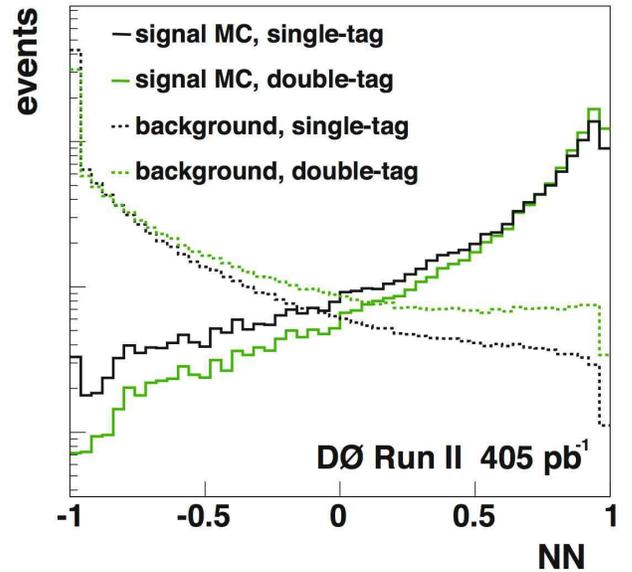}
\caption{\label{fig7} The output discriminant of an artificial neural network ($NN$) with six input nodes. All distributions are normalized to area.  $NN$ is optimized to distinguish between fully hadronic $t\bar{t}$ Monte Carlo events (signal) and the background from multijet production (background) as predicted by the tag rate functions. }
\end{figure}

The top quark production cross section was calculated from the output of {\em NN}, an artificial neural network trained to force its output near 1 for $t\bar{t}$ events and near $-1$ for QCD multijet events, using the multilayer perceptron in the {\sc root} analysis program\ \cite{root}. The six parameters illustrated in Figs.\ \ref{fig1}-\ref{fig6} were used as input for the neural net.
The very large background-to-signal ratio in the untagged data allowed us to use untagged data as background input for the training of {\em NN}, while $t\bar{t}$ MC was used for the signal.
Fig.\ \ref{fig7} shows the {\em NN} discriminant for $t\bar{t}$ signal and multijet background. 
Although the distributions for single- and double-tag events were different due to increased heavy flavor content in the double-tag sample, both samples showed a clear discrimination between signal and background.

\begin{figure}
\includegraphics[width=\linewidth]{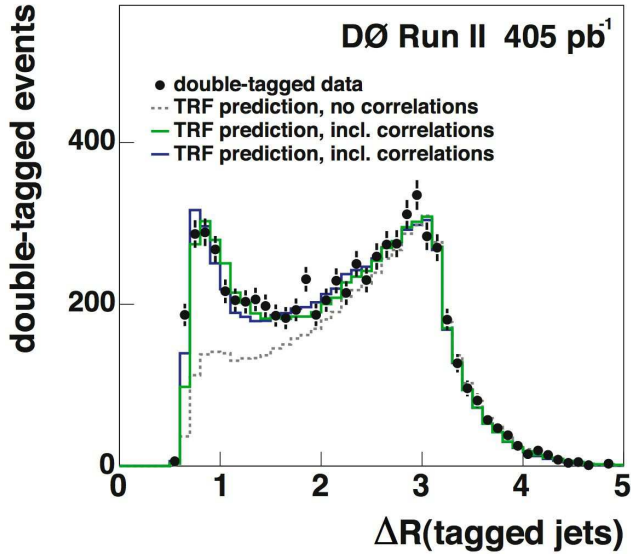}
\caption{\label{fig8} The performance of the {\sc trf} prediction on double-tag events (points), without including the correlation factor $C_{ij}$ (dashed histogram), and including $C_{ij}$ for two different functional parameterizations (solid histograms).}
\end{figure}

The overwhelming background also made it possible to use the entire (tagged and untagged) sample to estimate the background. 
For the loose and tight SVT, we derived a tag rate function ({\sc trf} --- the probability for any individual jet to have a secondary vertex tag ) from the data with $N_{tags} \leq 1$. 
The {\sc trf} was parameterized in terms of the $p_T$, $\phi$ and $y$ of the jet and the coordinate along the beam axis ($z$) of the primary vertex of the event, $z_{PV}$, in four different $H_T$ bins.
To predict the number of tagged jets in the event, it was necessary to correct for a possible correlation between tagged jets.
In the single-tag analysis the correlation factor was negligible, unlike in the double-tag analysis, where the presence of $b\bar{b}$+jets events in the sample enhanced the correlation correction.
We corrected for correlations caused by $b\bar{b}$ background by applying a correlation factor $C_{ij}$, that was parameterized as a function of the cone distance between the tagged jets, $\Delta R$. Figure\ \ref{fig8} shows the number of double-tagged events versus $\Delta R$ as observed in data, and the distribution as modeled by the {\sc trf} with and without including $C_{ij}$. 
We considered significantly different functional forms for the parameterization of $C_{ij}$ and found that the choice of parameterization had little effect on the shape of the modeled background distribution.

The probabilities $p_i$ were used to assign a weight, the probability that the event could have a given number of tags, to every tagged and untagged event in the sample.
To ensure the {\sc trf} prediction was accurate in the region of phase space outside the ``background'' peak of the neural network, we used the region $-0.7<NN<0.5$ to determine a normalization. 
In this region of phase space, the $t\bar{t}$ content was negligible.  A possible dependence on $t\bar{t}$ content was studied by the addition and/or subtraction of simulated $t\bar{t}$ events, as was the variation of the interval used for the normalization.
Outside the background peak, the {\sc trf} predictions were corrected by: {\em SF}$_{1} = 1.000 \pm 0.009$ for the single-tag analysis, and {\em SF}$_{2} = 0.969 \pm 0.014 $ for the double-tag analysis. The errors on the normalization were taken into account as a systematic uncertainty on the number of background events.

\begin{figure}
\includegraphics[width=\linewidth]{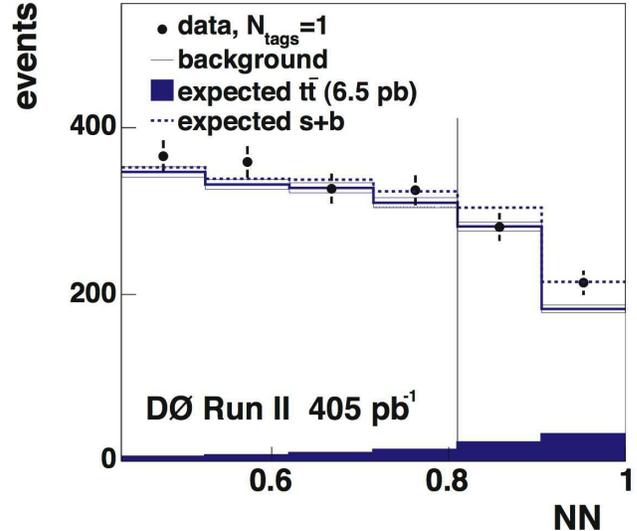}
\caption{\label{fig9} The distribution of the $NN$ output variable for single-tag events. Shown are the data (points), background (hashed band), signal (filled histogram) and signal+background (dashed histogram). The vertical line represents the used cut of $NN>0.81$.}
\end{figure}

\begin{figure}
\includegraphics[width=\linewidth]{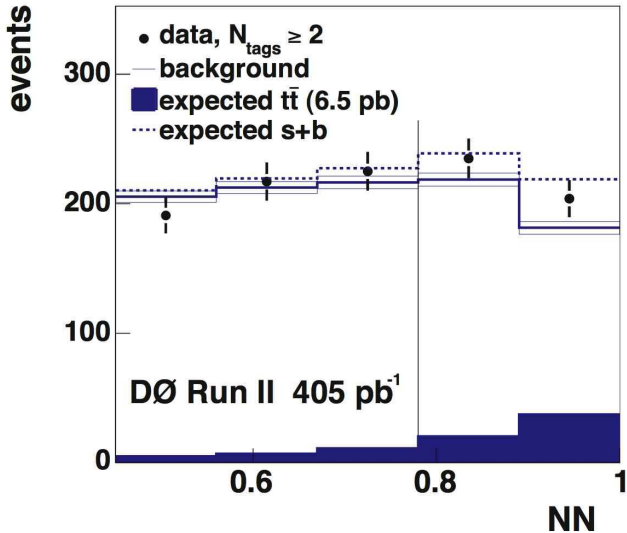}
\caption{\label{fig10} The distribution of the $NN$ output variable for double-tag events. Shown are the data (points), background (hashed band), signal (filled histogram) and signal+background (dashed histogram). The vertical line represents the used cut of $NN>0.78$. }
\end{figure}

Both the single-tag and the double-tag analysis were expected to be dominated by background, even at large values of $NN$. Figures\ \ref{fig9}  and\ \ref{fig10} show the distribution for data (points), the Monte Carlo simulation prediction for $\sigma_{t\bar{t}}=6.5~{\rm pb}$ (filled histogram), the background prediction (line histogram) and the signal+background distribution (dashed histogram)\ \cite{ttbarlepjetsvtxtag,theoryxsec}. 

The cross section was calculated from the number of $t\bar{t}$ and background candidates above a cut value of the $NN$ discriminant. 
The cut value was chosen to maximize the expected statistical significance $s/\sqrt{s+b}$, where $s$ and $b$ were the number of expected signal and background events.
The signal and background distributions were estimated using the {\sc trf} prediction and $t\bar{t}$ Monte Carlo events\ \cite{footnote3}. 
For both analyses, the expected statistical significance was about two standard deviations.  The optimal cut for the single (double)-tag analysis was $NN\geq 0.81~(0.78)$ shown by a vertical line in Figs.\ \ref{fig9}  and\ \ref{fig10}.
Table\ \ref{tabres1} gives the observed numbers of events ($N_{obs}^i$), the background prediction ($N_{bg}^i$) and the efficiency for signal ($\varepsilon_{t\bar{t}}$) that can be used to calculate the $t\bar{t}$ production cross section via:
\begin{equation}
\sigma_{t\bar{t}}=\frac{N_{obs}^i - N_{bg}^i}{ \varepsilon_{t\bar{t}}^i \mathcal{L} (1 - \varepsilon_{TRF}^i) } ,
\end{equation}
where $i$ was ``$=1$'' for the single-tag analysis and ``$\geq 2$'' for the double-tag analysis. 
The number of background events is predicted using the {\sc trf} method. It was likely that at values of $NN$ close to unity a certain fraction of the sample used to predict the background actually consists of tagged or untagged $t\bar{t}$ events, resulting in an increased background prediction.
The expected $t\bar{t}$ contamination of the background sample was corrected by a factor $\varepsilon_{TRF}^i$. In the higher value bins of $NN$, the contribution from untagged $t\bar{t}$ events was significant. $\varepsilon_{TRF}^i$ was estimated by applying the {\sc trf} on $t\bar{t}$ MC, and comparing the predicted tagging probability for signal to what was expected from background. The size of the Monte Carlo sample dominates the uncertainty on  $\varepsilon_{TRF}^i$.  

Table\ \ref{tabres1} lists the systematic uncertainties on the estimate of the number of background events, the selection efficiency and the background contamination. The first was uncorrelated between the two analyses, while the latter two were correlated as they were derived from the same Monte Carlo samples. 

For the single-tag analysis, the systematic uncertainty on the selection efficiency was dominated by the uncertainty in the jet calibration and identification, which were estimated by varying the parameterizations used by one standard deviation.  
The uncertainty on the background prediction was dominated by the uncertainty on the {\sc trf} method and the uncertainty on $\varepsilon_{TRF}$ was due to limited Monte Carlo statistics. 
The uncertainty of the {\sc trf} prediction was comprised from the uncertainties coming from the fits of the probability density functions at the jet level, the statistics of the background sample and the uncertainty on the normalization and correlation factors $SF$ and $C_{ij}$. 
For the double-tag analysis, the contribution from the uncertainties due to calibration of the $b$ quark jet identification efficiency was an additional systematic uncertainty on $\varepsilon_{t\bar{t}}$. 
These uncertainties were derived by varying the parameterizations used within their known uncertainties.

\begin{table}
\caption{\label{tabres1}Overview of observed events, background predictions and efficiencies.}
\begin{tabular}{l c c c}
\hline \hline
  ~ & symbol  & value \\ \hline
observed events& $N_{obs}^{=1}$   & 495  \\
background events& $N_{bg}^{=1}$  & $464.3 \pm 4.6 ({\rm syst})$ \\
$t\bar{t}$ efficiency & $\varepsilon_{t\bar{t}}^{=1}$    & $0.0242 _{-0.0058}^{+0.0049}({\rm syst})$ \\
$t\bar{t}$ contamination & $\varepsilon_{TRF}^{=1}$   & $0.245 \pm 0.031 ({\rm syst})$ \\  \hline
observed events & $N_{obs}^{\geq 2}$  &  439 \\
background events& $N_{bg}^{\geq 2}$  &  $400.2_{-6.2}^{+7.3} ({\rm syst})$  \\ 
$t\bar{t}$ efficiency & $\varepsilon_{t\bar{t}}^{\geq 2}$   & $0.0254 _{- 0.0070}^{+0.0065} ({\rm syst})$ \\
$t\bar{t}$ contamination & $\varepsilon_{TRF}^{\geq 2}$    & $0.194 \pm 0.048({\rm syst})$
 \\ 
\hline \hline
\end{tabular}
\end{table}

The single-tag analysis yielded a cross section of 
\begin{equation}
\sigma_{t\bar{t}}=4.1_{-3.0}^{+3.0}({\rm stat}) _{-0.9}^{+1.3}({\rm syst}) \pm 0.3 ({\rm lumi})~{\rm pb}. 
\end{equation}
For the double-tag analysis the measured cross section was 
\begin{equation}
\sigma_{t\bar{t}}=4.7_{-2.5}^{+2.6}({\rm stat}) _{-1.4}^{+1.7}({\rm syst}) \pm 0.3 ({\rm lumi})~{\rm pb}. 
\end{equation} 
As the single-tag and double-tag analysis were measured on independent samples, the statistical uncertainties were uncorrelated. 
The uncertainties on the selection efficiency were completely correlated. 
Taking all uncertainties into account, a combined cross section measurement of 
\begin{equation}
\sigma_{t\bar{t}}=4.5_{-1.9}^{+2.0}({\rm stat}) _{-1.1}^{+1.4}({\rm syst}) \pm 0.3 ({\rm lumi})~{\rm pb} 
\end{equation}
was obtained, for a top quark mass of $m_t=175~{\rm GeV}/c^2$. 
For a top quark mass of $m_t=165~{\rm GeV}/c^2$, the cross section is $\sigma_{t\bar{t}}(165)=6.2_{-2.7}^{+2.8}({\rm stat}) _{-1.5}^{+2.0}({\rm syst}) \pm 0.4 ({\rm lumi})$ pb, while for a top quark mass of $m_t=185~{\rm GeV}/c^2$ the value shifted down to $\sigma_{t\bar{t}}(185)=4.3_{-1.8}^{+1.9}({\rm stat}) _{-1.0}^{+1.4}({\rm syst}) \pm 0.3 ({\rm lumi})$ pb.

In summary, we have measured the $t\bar{t}$ production cross section in $p\bar{p}$ interactions at $\sqrt{s}=1.96$ TeV in the fully hadronic decay channel. 
We used lifetime $b$-tagging and an artificial neural network to distinguish $t\bar{t}$ from background. 
Our measurement yields a value consistent with SM predictions and previous measurements.

%
We thank the staffs at Fermilab and collaborating institutions, 
and acknowledge support from the 
DOE and NSF (USA);
CEA and CNRS/IN2P3 (France);
FASI, Rosatom and RFBR (Russia);
CAPES, CNPq, FAPERJ, FAPESP and FUNDUNESP (Brazil);
DAE and DST (India);
Colciencias (Colombia);
CONACyT (Mexico);
KRF and KOSEF (Korea);
CONICET and UBACyT (Argentina);
FOM (The Netherlands);
PPARC (United Kingdom);
MSMT (Czech Republic);
CRC Program, CFI, NSERC and WestGrid Project (Canada);
BMBF and DFG (Germany);
SFI (Ireland);
The Swedish Research Council (Sweden);
Research Corporation;
Alexander von Humboldt Foundation;
and the Marie Curie Program.


\begin{thebibliography}{99}
%
\bibitem[*]{alton}
Visitor from Augustana College, Sioux Falls, SD, USA
\bibitem[\dag]{voutilainen}
Visitor from Helsinki Institute of Physics, Helsinki, Finland.
%
\vskip 0.25cm
  
 \bibitem{d0det}
D\O\ Collaboration, V.M. Abazov {\sl et al.}, Nucl. Instrum. Methods Phys. Res. A {\bf 565}, 463  (2006).
\bibitem{newlumi}
T.~Andeen {\sl et. al.}, FERMILAB-TM-2365-E (2006), in preparation.
\bibitem{footnote1}
The efficiency for signal remained between 85 and 90\% throughout the data collection period. Efficiencies were was measured both on $t\bar{t}$ Monte Carlo and derived from parameterizations determined from data.
\bibitem{alpgen}
M.L. Mangano {\sl et al.}, J. High Energy Phys. {\bf 07}, 001 (2003).
\bibitem{pythia}
T. Sj\"{o}strand {\sl et al.}, Comput. Phys. Commun. {\bf 135}, 238 (2001).

\bibitem{evtgen}
D. Lange, Nucl. Instrum. Methods Phys. Res. A {\bf 462}, 152 (2001).
\bibitem{jetsdef}
G.C. Blazey {\sl et al.}, 
in {\sl Proceedings of the Workshop: QCD and Weak Boson Physics in Run II}, 
U. Baur, R.K. Ellis and D. Zeppenfeld (ed.), Fermilab, Batavia, IL (2000).
 \bibitem{ttbarlepjetsrun2}
D\O\ Collaboration, V.M. Abazov {\sl et al.}, 
Phys. Lett. B {\bf 626}, { 35} (2005).
\bibitem{ttbarlepjetsvtxtag}
D\O\ Collaboration, V.M. Abazov {\sl et al.}, 
Submitted to Phys. Rev. D, FERMILAB-PUB-06-386-E (2006).
\bibitem{pdg}
W.-M.~Yao {\sl et al.}, Journal of Physics G {\bf 33}, 1 (2006).
\bibitem{run1alljetsxsec1}
D\O\ Collaboration, B. Abbott {\sl et al.}, 
Phys. Rev. Lett. {\bf 83} 1908 (1999).
\bibitem{footnote2}
The possibility that the wrong permutations of jets could be chosen was taken into account in the determination of the values of the values of 79, 11, and 21 GeV/$c^2$ for $M_W$, $\sigma_{M_W}$ and $\sigma_{m_t}$.
\bibitem{root}
R. Brun and F. Rademakers, 
Nucl. Inst. Meth. in Phys. Res. A {\bf 389} (1997) 81-86. See also http://root.cern.ch/.

\bibitem{theoryxsec}
N. Kidonakis and R. Vogt, Phys. Rev. D {\bf 68}, 114014 (2003).
\bibitem{footnote3}
The expected $t\bar{t}$ content used to optimize the $NN$ cut was equivalent with a hypothetical cross section of $\sigma_{t\bar{t}}=6.5~{\rm pb}$. The chosen cuts are stable under variation of the value assumed for the optimization.
\end{thebibliography}
\end{document}